\documentclass[reprint,graphicx,amsmath,amssymb,]{revtex4-1} %% use 12pt for preprint option
\usepackage{amsmath,amssymb,graphicx,xcolor}
\usepackage{graphicx}% Include figure files
\usepackage{dcolumn}% Align table columns on decimal point
\usepackage{bm}% bold math
\usepackage{natbib}
\usepackage{acronym}
\usepackage{float}
\newcommand{\be}{\begin{equation}}
\newcommand{\ee}{\end{equation}}

\newcommand{\vdiff}{\Delta V}

\newcommand{\vpump}{\nu_{\mathrm{pump}}}
\newcommand{\vrepump}{\nu_{\mathrm{repump}}}

\newcommand{\alop}{\alpha_0^{(P)}}
\newcommand{\alos}{\alpha_0^{(S)}}
\newcommand{\altp}{\alpha_2^{(P)}}

\newcommand{\eglass}{\epsilon_{\mathrm{glass}}}

\begin{document}

\title{Transparent Electrodes for High E-Field Production Using A Buried ITO Layer}

\author{Will Gunton}
\affiliation{
Department of Physics and Astronomy, University of British Columbia, 6224 Agricultural Road, Vancouver, British Columbia, Canada V6T 1Z1}
\author{Gene Polovy}
\affiliation{
Department of Physics and Astronomy, University of British Columbia, 6224 Agricultural Road, Vancouver, British Columbia, Canada V6T 1Z1}
\author{Mariusz Semczuk}
\affiliation{
Department of Physics and Astronomy, University of British Columbia, 6224 Agricultural Road, Vancouver, British Columbia, Canada V6T 1Z1}
%\affiliation{Institute for Quantum Optics and Quantum Information, Austrian Academy of Sciences, Boltzmanngasse 3, Vienna A-1090, Austria}
\author{Kirk W.~Madison}
\affiliation{
Department of Physics and Astronomy, University of British Columbia, 6224 Agricultural Road, Vancouver, British Columbia, Canada V6T 1Z1}

\date{\today}% It is always \today, today,
             %  but any date may be explicitly specified

\begin{abstract}
We present a design and characterization of optically transparent electrodes suitable for atomic and molecular physics experiments where high optical access is required.  The electrodes can be operated in air at standard atmospheric pressure and do not suffer electrical breakdown even for electric fields far exceeding the dielectric breakdown of air.  This is achieved by putting an ITO coated dielectric substrate inside a stack of dielectric substrates, which prevents ion avalanche resulting from Townsend discharge. With this design, we observe no arcing for fields of up to 120 kV/cm.  Using these plates, we directly verify the production of electric fields up to 18~kV/cm inside a quartz vacuum cell by a spectroscopic measurement of the dc Stark shift of the $5^2S_{1/2} \rightarrow 5^2P_{3/2}$ transition for a cloud of laser cooled Rubidium atoms.  We also report on the shielding of the electric field and residual electric fields that persist within the vacuum cell once the electrodes are discharged. In addition, we discuss observed atom loss that results from the motion of free charges within the vacuum.  The observed asymmetry of these phenomena on the bias of the electrodes suggests that field emission of electrons within the vacuum is primarily responsible for these effects and may indicate a way of mitigating them.
\end{abstract}

%\ocis{(140.3490) Lasers, distributed-feedback; (060.2420) Fibers, polarization-maintaining; (060.3735) Fiber
%Bragg gratings; (060.2370) Fiber optics sensors.}% REPLACE WITH CORRECT OCIS CODES FOR YOUR ARTICLE
%                          % NOTE: \ocis{} IS ALIASED TO \pacs{} BUT MUST
%                          % FORMAT THE TERMS CORRECTLY FOR EACH JOURNAL

\maketitle %% required

While the application and use of large dc magnetic fields is common in atomic and molecular experiments, many would benefit from, or even require, the application of moderate to high dc electric fields (greater than 5~kV/cm). Examples include experiments with polar molecules \cite{Ni10102008, B911779B, DipolarSpinExchange} where large fields (greater than 10~kV/cm for the alkali dimers) are required to achieve partial lab frame alignment and more than 50\% of the maximum dipole moment. In these systems, electric fields can be used to, for example, measure the permanent electric dipole moment of the molecules \cite{B911779B} and control their total collisional cross section \cite{PhysRevA.76.052703}. Moderate to large electric fields are also required in experimental searches for the electron or neutron electric dipole moment \cite{Hudson, Regan}, controlling the dynamics of electronic spin relaxation of cold molecules \cite{Tscherbul}, or creating metastable weakly bound pairs (``field linked'' states) of ultra-cold polar molecules, whose properties depend strongly on the strength of the field \cite{PhysRevLett.90.043006,PhysRevA.69.012710}. 

Other proposed experiments involve measurements of the effect of electric fields on atomic and molecular collision resonances \cite{Krems,LiZ, PhysRevA.79.042711}.  For example, studying the effect of electric fields on heteronuclear Feshbach resonances in Li+Rb mixtures requires electric fields in excess of 20~kV/cm, with even higher fields needed for less polar mixtures like K+Rb \cite{K.-K.Ni10102008}, Rb+Cs \cite{PhysRevLett.113.205301,PhysRevLett.113.255301}, or Na+K \cite{PhysRevLett.114.205302}. This was the primary motivation for the creation of the electric field plates discussed in this paper.

Many AMO experiments involving the application of electric fields share the same requirements - namely, electrodes that do not limit optical access and that have extremely low residual magnetism.  Thin transparent conducting films (TCFs), such as indium tin oxide (ITO), applied to a transparent dielectric substrate satisfy these requirements. The ability to place the electrodes outside of a vacuum chamber allows for the addition of electric field capability after the design or construction of the apparatus. In addition, placing the electrodes in air simplifies the vacuum design, and can increase the flexibility of an experiment (most obviously because they can be added and removed when necessary).

The primary limitation to creating large electric fields is dielectric breakdown. This phenomenon occurs when free electrons are accelerated to sufficiently high energies by an electric field such that they impact ionize molecules or atoms in the surrounding medium or in the electrode itself. This leads to an avalanche production of many more electrons and ions (by electron impact ablation) generating an electric arc that rapidly reduces the field.  Dielectric breakdown limits the maximum electric field that can be generated when the flow of charge through the medium exceeds the current supplied by the high voltage source. Breakdown also leads to material damage of the electrodes and other surfaces due to ablation and X-ray generation through bremsstrahlung which can also damage materials and sensitive electronics.

\begin{figure}[htbp]
\includegraphics[width=0.46\textwidth]{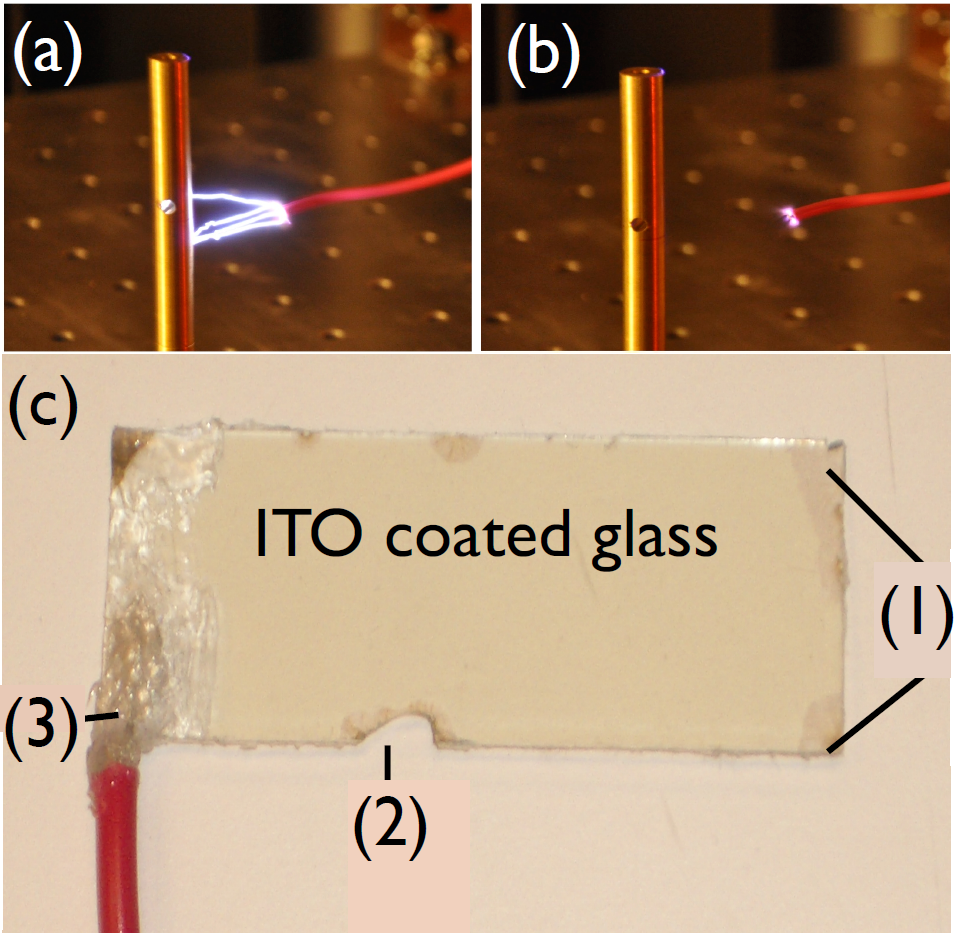}
\caption{(color online) Illustration of the effects of Townsend discharge and ion avalanche.  In (a) an electric arc is established due to the breakdown of air across a 2 cm gap between a grounded metal rod and the tip of an exposed wire near 45 kV.  In (b) the output current of the high voltage supply is limited and the electric arc is not sustained.  Nevertheless, the faint blue glow of a corona discharge near the tip of the wire is visible indicating field electron emission.  In (c), an ITO coated slide is shown, and the deleterious effects of discharge are apparent.  At (1) the ITO coating appears darker and was damaged by corona discharge at the corners where the local electric field is highest.  At (2), the glass cracked and broke due to electrical arcing.  Conductive epoxy (3) was used to glue and electrically connect the conductor to the ITO coating and was insulated using RTV silicone adhesive.
}
\label{fig:breakdown}
\end{figure}

Even when complete dielectric breakdown does not happen (and an electric arc is not present), discharge can still occur and is often evident from a corona discharge occurring near sharp edges (see Fig.~\ref{fig:breakdown}).  The large curvature associated with an edge of a conductor causes a large potential gradient (i.e.~a large local electric field) which can lead to field electron emission followed by a local dielectric breakdown.  Field emission and corona formation can be suppressed by designing electrodes without sharp edges (i.e.~by using large-diameter, round-shaped conductors) and by polishing the surface to remove microscopic surface roughness \footnote{Polishing of electrode surfaces to remove microscopic roughness can be achieved by controlled discharge where the supply current is strongly limited. The polishing occurs by local ion ablation of the surface generated by the field electron emission of the rough feature itself in the presence of a gas (e.g. Argon)}.

Optically transparent electrodes are important elements in a variety of technologies (including liquid-crystal displays, touchscreens and photovoltaics) and are created by the application of a TCF to a transparent and electrically insulating substrate.  TCFs are electrically conducting but only optically transparent in thin layers.  ITO is the most common TCF, but other transparent conductive oxides exist, and other alternatives exist for thin conductive films including conductive polymers, films of carbon nanotubes, graphene, and metallic nanostructures \cite{ADMA:ADMA201003188}.  One obvious challenge to creating large electric fields with a TCF is that electric discharge and the ion ablation that accompanies it can easily destroy a thin film (see Fig.~\ref{fig:breakdown}c).  Thus suppressing discharge of any kind is of key importance.

\section{Previous work}

Transparent electrodes capable of creating moderately high fields (on the order of 5~kV/cm) for experiments with ultra-cold molecules have been realized with ITO coated glass (\#CH-50IN-S209 from Delta Technologies) by the group of Ye and Jin \cite{Ni10102008, B911779B, DipolarSpinExchange}. The ITO glass was left exposed and the dielectric breakdown and discharge from the plates was limited by covering all nearby conductors with many layers (4-5) of Kapton tape. With this system, they claim to be able to apply up to $\pm 5$~kV on each plate without breakdown generating a 9~kV/cm field with a plate separation of 1.36~cm. They observed that for applied voltages of $\pm 3$~kV on each plate (corresponding to 5.2~kV/cm field) a residual electric field would remain after the plates were discharged \cite{Ni-Thesis}. They conjectured that their pyrex glass cell (made from Borofloat by Starna Cells) developed a residual electric polarization.

Other groups working with polar molecules have produced moderate electric fields of up to 2 kV/cm using either ITO slides \cite{PhysRevLett.114.205302} or using four parallel rods at the corners of the vacuum cell \cite{PhysRevLett.113.205301}.

%Dielectric breakdown and the discharge of their plates was limited by covering nearby conductors with many layers of Kapton tape.
%They tested the breakdown of the tape by putting a grounded wire wrapped in 2 layers of Kapton tape directly on the coated surface.  They observed discharge at 4~kV.
%With this system, they claim to be able to apply up to $\pm 5$~kV on each plate without breakdown generating a 9~kV/cm field for a plate separation of 1.36~cm.However, they observed that for applied voltages of $\pm 3$~kV on each plate (corresponding to 5.2 kV/cm field) a residual electric field would remain after the plates were discharged \cite{Ni-Thesis}.
%They conjectured that their pyrex glass cell (made from Borofloat by Starna Cells) developed a residual electric polarization.

\section{This design}

\begin{figure}[htbp]
\includegraphics[width=0.46\textwidth]{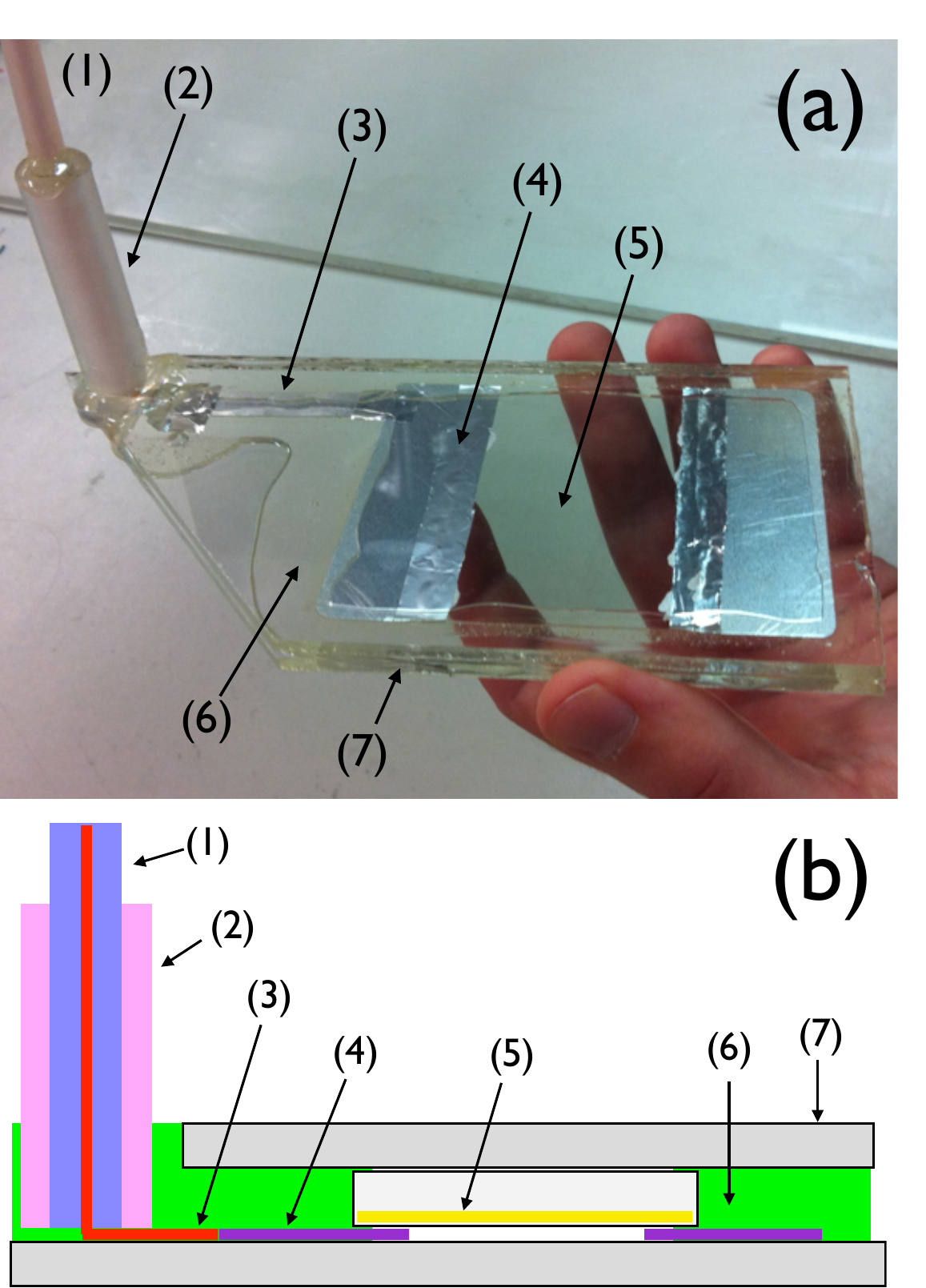}
\caption{(color online) (a) Picture and (b) schematic (not to scale) of the electric field plate assembly.  A high voltage coaxial cable (polyethylene RG-8U) with a 20~cm length stripped of its ground shield (1) is inserted into a 10~cm cylindrical tube of either Teflon or, here, acylic (2) that extends down to the lower glass plate, and the center conductor connects to a flat metal conductor (3) that extends to aluminum foil strips (4) that contact the ITO coated glass slab (5).  Insulating epoxy (6) glues the heterostructure together, maintains the positions of all elements, and fills the gap between the external glass plates (7).
}
\label{fig:field_plate_image}
\end{figure}

In our work, a TCF of ITO was also used. However, electrical breakdown for applied voltages of up to 60~kV is prevented by embedding the ITO coated dielectric substrate inside of a stack of two more transparent substrates (making a glass sandwich) where the outer layers have a much higher dielectric breakdown.  Fig.~\ref{fig:field_plate_image} shows a picture and schematic of the whole assembly.  An ITO coated glass slide from SPI Supplies (5~cm $\times$ 7.5~cm $\times$ 0.7~mm) was laid onto a borosilicate glass flat (5~mm thick) with the coating side down and partially overlapping two strips of aluminum foil, which extend the conductive layer area to a 7.5~cm $\times$ 9.5~cm area. In a first iteration of the plates, we used a conductive surface that measured just under 3~cm $\times$ 3~cm. However, we elected to build the final version of the plates with a much larger conductive surface because of the scaling of the magnitude of the electric field between the plates with the plate size. In fact, the larger plates have an expected electric field strength (at a plate separation of 4~cm) that is 50\% larger than the smaller design at the same applied voltage. The final size was chosen in order to maximize the electric field strength and homogeneity inside the cell, given the limitations imposed by other elements in the apparatus.

We applied 5-minute epoxy made by Devcon to the area just outside of the ITO slab up to the edge of the bottom glass plate and another glass plate was pressed to the top to complete the stack.  At least a 1~cm border filled with epoxy existed between the conductor and the edge of the glass sandwich. This ensured that a dielectric breakdown would not occur through the epoxy. A long flat conductor embedded in the sandwich connects the foil strip to the center conductor of a high voltage coaxial cable (polyethylene RG-8U).  The cable's connection to the plate was found to be a weak spot for breakdown, and care was taken to create a rigid and highly insulated connection.  To make this connection, the end of the high voltage coaxial cable was stripped of its ground shield and inserted into a 10~cm long cylindrical tube made of either Teflon or acrylic (both were found to be suitable although we found it important to roughen the Teflon surface before gluing to improve the bond) with a wall thickness of 5~mm and that extended down to the lower glass plate.   We used 5-minute epoxy to affix the tube to the glass, and to fill the inside of the tube to act as electrical insulation for the exposed wire.  To further secure the assembly, rubber mastic tape (.065 thickness from 3M/Scotch) was wrapped around the tube and the areas where the tube and glass assembly contacted (not shown in Fig.~\ref{fig:field_plate_image})

%This teflon cylinder provided a rigid and highly insulated connection that prevented breakdown from occurring at the point where the inner conductor is connected to the flat conductor inside the sandwich.

\begin{figure}[h]
\centering
\includegraphics[width=0.2\textwidth]{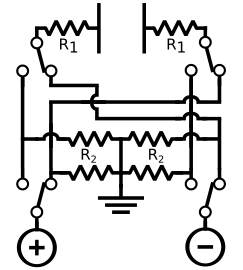}
\caption{Schematic for high voltage switching network for the electrodes. The $R_1 = 4~$M$\Omega$ resistor in series with the electric field plates helps limit the instantaneous current spikes during the charging and discharging of the plates, while the $R_2 = 1~$G$\Omega$ resistor in each connection arm ensures that all non-operational arms are well grounded, and allows for a rapid discharge of the plates when the supply is disconnected. The high voltage relays are controlled via an optically isolated digital switch.}
\label{fig:schematic}
\end{figure}

A high voltage switching network (similar to an H bridge) was built for controlling the polarity of the voltages applied to the plates. The network is constructed using four single pole double throw Gigavac high voltage relays (G71C771) and two Glassman 60~kV supplies (EH60R01.5), as shown in Fig.~\ref{fig:schematic}. Four $1~$M$\Omega$ Ohmite ceramic resistors (OY105KE) were placed in series between the high voltage supplies and the electric field plates in order to limit the current spikes that occur during charge and discharging. In addition, each path from the voltage supplies to either of the electric field plates is connected to ground through a 75~kV, 16~W, $1~$G$\Omega$ Ohmite high voltage resistor (MOX96021007FTE). These resistors ensure that all non-operational arms of the network are grounded and allow a path for a rapid discharge of electric field plates.

In addition to the electric plates, care had to be taken to insulate all electrical connections including those to resistors or those between lengths of cable. This was achieved by embedding all connections inside of extruded 1 inch diameter acrylic rods or blocks. To make the connections, a through hole was drilled along the axis of the rod through which the two cables were inserted. A perpendicular access hole was drilled, and used to access the connection for soldering. Care was taken to ensure that solder connections had no sharp edges, and Super Corona Dope (from MG Chemicals) was applied to the connection to help insulate and resist corona formation. Finally, all the holes were filled with 5-minute epoxy or melted resin wax.  We found that filling volumes through narrow channels was more easily done with wax than the viscous epoxy.  Due to the lack of availability of commercial high voltage connectors, almost all of the high voltage connections used in this apparatus were permanent and hand-made. 

\section{Plate testing}

Using the design presented above, we observed that no arcing would occur when a grounded conductor was placed touching any part of the assembly with the plate energized to either $\pm 60$~kV.  Given the minimum distance from the inner conductor surfaces to the outside of the assembly was 5~mm, this corresponded to a maximum electric field strength sustained across the outer glass substrate of 120~kV/cm.  When the electric field plates were operated in the laser cooling setup, we placed a grounded cable next to the connection between the high voltage cable and the plate to ensure that, in the event of a failure of this component, the arc would not be to the optical table or to other equipment in the apparatus.

During testing of the switching network, we noted that charging and discharging the plates occasionally had detrimental effects on electronic equipment close to the switching network. We attributed this to transient spikes and/or radiation produced during the charging period. To avoid these effects on our experimental electronics, the switching network was located 10~m from the experimental setup in an adjacent room, and the high voltage relays were controlled via optically isolated digital switches.

\section{Electric field measurements}
\label{sec:efm}

\begin{figure}[h]
\centering
\includegraphics[width=0.46\textwidth]{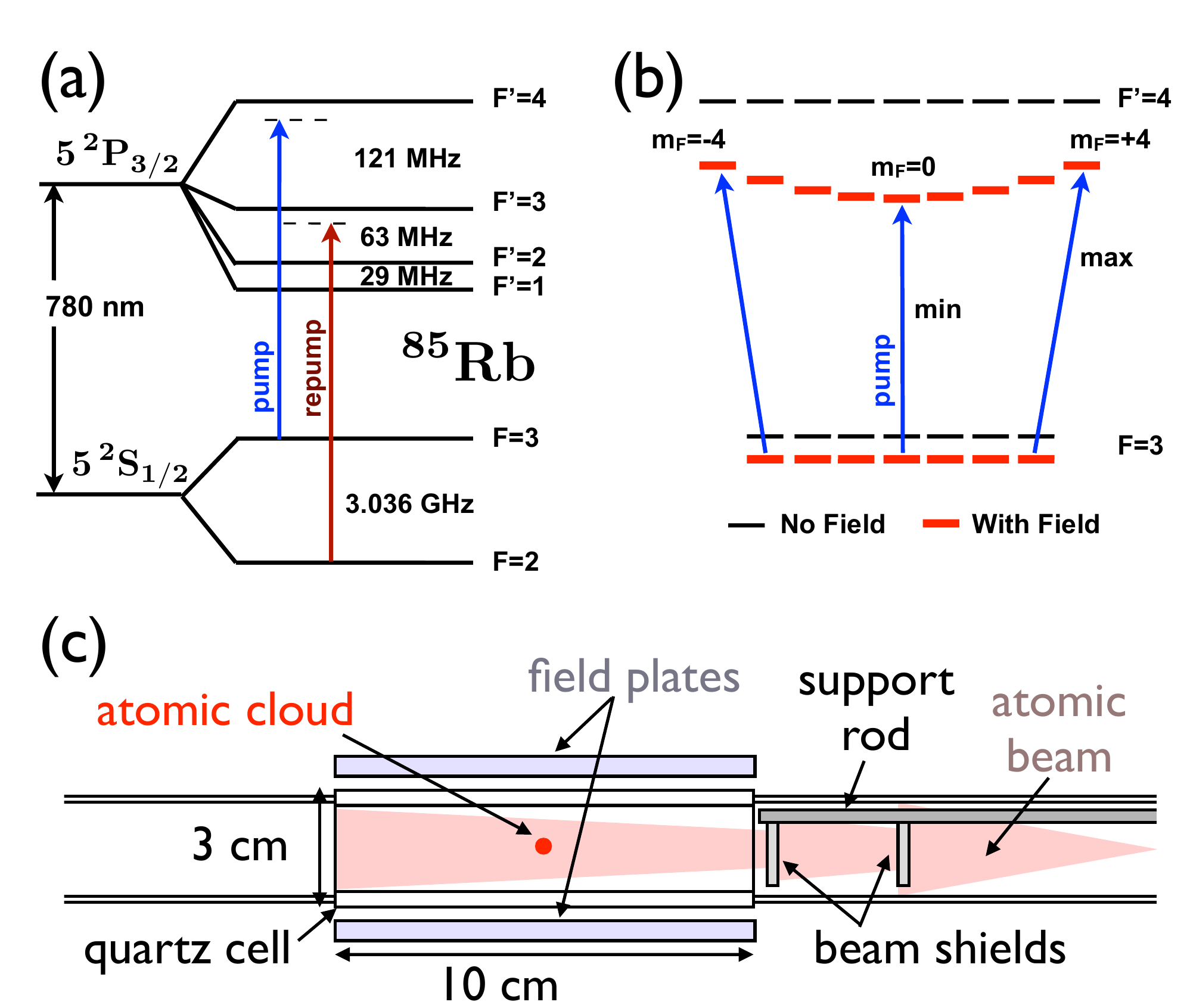}
\caption{(a) Atomic energy levels (not to scale) and transitions used for magneto-optic trapping of $^{85}$Rb atoms.  In (b) the dc Stark shift of the ground ($F=3$) and excited state ($F'=4$) levels is illustrated.  The scalar polarizability is larger in the excited state and it therefore shifts down more than the ground state.  The tensor polarizability (zero for the ground state) shifts the excited state levels up by an amount proportional to $m_F^2$.  Thus the pump transition moves to lower frequencies and also broadens somewhat indicated by the ``min" and ``max" transition energy differences. In (c) the placement of the field plates on a quartz vacuum cell is shown for the generation of a high electric field imposed on a sample of Rb atoms held in a magneto-optic trap.  A grounded support rod holding the atomic beam shields was less than 1~cm from the top plate and is, unfortunately, a source of electrons by field electron emission when the top plate is positively biased.
}
\label{fig:TrappingLightStarkShift}
\end{figure}

Panel c of Fig.~\ref{fig:TrappingLightStarkShift} shows how we incorporated two of these field plates into our laser cooling apparatus described in \cite{Ladouceur:09,PhysRevA.88.023624}. Although we experienced a loss of power due to reflections from the transparent substrates (which were not AR coated) and the ITO substrate (the transmission around 670~nm and 780~nm is about 80\%), we did not experience any significant decrease in the MOT performance. While Rb atoms were introduced into the vacuum with an atomic dispenser placed 30~cm from the trapping region, Li atoms were supplied to the trapping region by an atomic beam.  In order to prevent the buildup of Li, emitted by the oven, within the main section of the chamber, a series of beam shields were used.  These shields were mounted on a metal support rod that ran along the top of the cell, as pictured.

Although the minimum field-plate conductor separation is limited to 4~cm by the fused silica vacuum cell (3~cm) and the outer glass plate on the electrode assembly (5~mm), because of the high dielectric constant of both glass and quartz (for example, $\eglass \sim 3$ to $5$), the field is larger than simply the voltage difference $\vdiff$ divided by the physical distance, $d$.  In fact, the strength of the electric field is
\be
|E| = \frac{\vdiff}{d}\cdot\frac{d}{d - t (1 - 1/\eglass)} = \frac{\vdiff}{d}\cdot\eta,
\ee
where $t$ is the total thickness of the glass and quartz layers, $d$ is the physical separation between the conducting layers and the dimensionless quantity $\eta$ is approximately 1.5.  Thus, by applying $\pm$~60~kV to each plate, we expected to obtain a field of approximately 45~kV/cm.

In order to measure the electric fields produced by these plates, we performed \emph{in situ} spectroscopy on a sample of laser cooled Rb atoms inside the vacuum cell with the plates energized.  Fig.~\ref{fig:TrappingLightStarkShift} shows the energy levels for $^{85}$Rb and both the ``pump" ($F=3 \rightarrow F=4'$) and ``repump" ($F=2 \rightarrow F=3'$) light tuned to the $5^2S_{1/2} \rightarrow 5^2P_{3/2}$ transition for  laser cooling and trapping of Rb.  Also shown is an illustration of the Stark effect on these levels for the pump transition.

The dc Stark interaction Hamiltonian is

\be
H_E = \frac{1}{2} \alpha_0 E_z^2 - \frac{1}{2} \alpha_2 E_z^2 \frac{3J_z^2 - J(J+1)}{J(2J-1)}
\ee

where the scalar polarizability for the ``P" excited state is larger than the ``S" ground state ($\alop > \alos$) and the tensor polarizability ($\altp < 0$) is only non-zero for the ``P" state (for $J=3/2$). For small electric fields where the shift is small compared to the hyperfine splittings, the energy difference between the ground and excited states is
\be
\Delta E =  \frac{1}{2} (\alop - \alos) E_z^2 - \frac{1}{2} \altp E_z^2 [3 m_F^2 - F(F+1)] \cdot k,
\ee
where 
\be
k = \frac{[3X(X-1) - 4F(F+1)J(J+1)]}{(2F+3)(2F+2)F(2F-1)J(2J-1)},
\ee
and $X=F(F+1) + J(J+1) - I(I+1)$ \cite{steck85}. Note $m_F$, $F$ and $J$ are the quantum numbers for the excited state.

The addition of an electric field thus shifts the ``pump" transition to lower frequencies and also broadens it by breaking the degeneracy of the $m_F$ levels in the excited state.

\begin{figure}[htbp]
\includegraphics[width=0.45\textwidth]{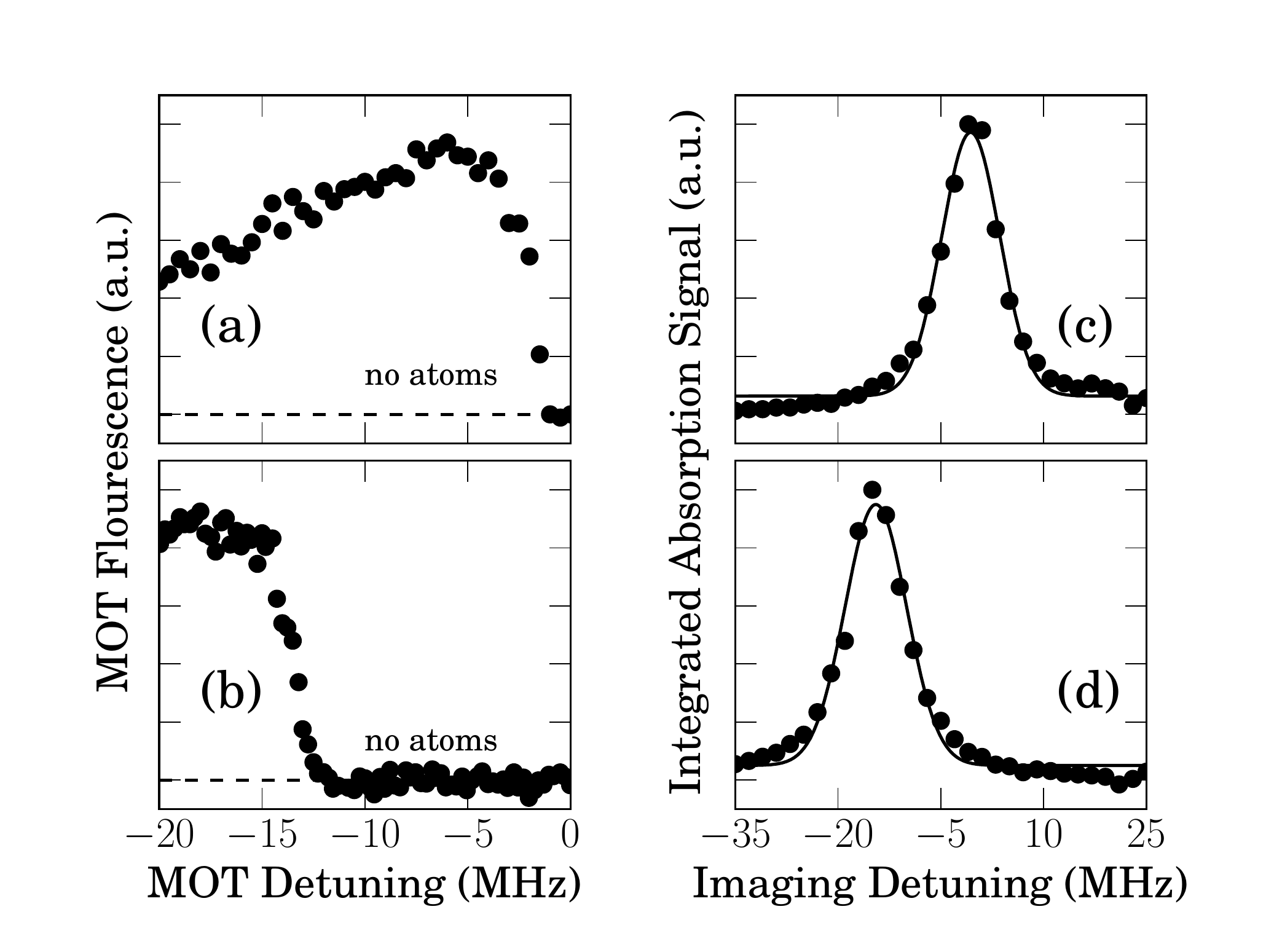}
\caption{The influence of the dc Stark effect on the operation of the $^{85}$Rb MOT (a and b) and on the integrated absorption of a laser beam by the cold $^{85}$Rb atomic cloud in the absence of a magnetic field versus detuning (c and d).  In (a) and (b), the MOT fluorescence is plotted versus the pump detuning.  With no electric field present, the MOT operates for detunings as small as -2~MHz, whereas this transition point shifts down by more than 11 MHz with a field of 14~kV/cm applied by the plates.  The dashed line indicates the fluorescence level consistent with no atoms trapped.  In (c) and (d) the integrated absorption of a imaging beam (with the MOT light extinguished) is plotted as a function of the detuning of the imaging light from resonance.  A repumping beam was simultaneously applied during the absorption measurement.  The frequency at which the absorption is maximum shifts down by $13.9$~MHz corresponding to the expected shift from the applied electric field of 14~kV/cm.  The solid line is a fit to the data where the mean value for the fits are $-0.6$~MHz in (c) and $-14.5$~MHz in (d).
}
\label{fig:StarkMeasurements}
\end{figure}

Two types of measurements were performed, and are shown in Fig.~\ref{fig:StarkMeasurements}.  In the first case, we measured the Rb MOT fluorescence as a function of the detuning of $\vpump$ from resonance with $\vrepump$ set on resonance.  Here the ``on resonance" values are defined in the absence of any Stark shift.  With no electric field present, the MOT is observed to operate at pump detunings as small as -2~MHz whereas with a 14~kV/cm applied field we observed the MOT operating only for pump detunings larger than -13~MHz.  An accurate determination of the electric field from the overall shift is confounded by the broadening of the operating point of the MOT and due the unknown residual Zeeman shift resulting from light pressure imbalances which push the MOT center into a region where the magnetic field is non-zero.

To get a more easily interpreted measure of the dc Stark shift at the center of the cell, we perform a magnetic-field-free measurement of the absorption of the atoms.  For this, we imaged the shadow cast by the Rb atomic cloud in a laser beam tuned to the ``pump" transition with the MOT light and magnetic field extinguished.  The absorption profile was integrated, and the total integrated signal is plotted as a function of the imaging light detuning.  The maximum absorption of the cloud was observed to shift down by $13.9$~MHz from the electric-field-free measurement corresponding to the expected average frequency shift from the applied electric field of 14~kV/cm.  As our state population in the $F=3$ ground level just before imaging is approximately equally distributed among the $m_F$ states, we use the mean Stark shift of the pump transition to determine the electric field strength.  Any associated uncertainty on the determination of the field strength is due to the unknown state populations. As a reference, if the spin population were all in the $|m_\mathrm{F}|=3$ states, we would underestimate the field by 20\% and if the population were all in the $m_\mathrm{F}=0$ we would overestimate the field by 10\%.

%The signal maximum location versus the expected field (based on the applied voltage and geometry) is plotted in the inset of Fig.~\ref{fig:StarkMeasurements}(d). The data appear to follow the mean Stark shift indicating that the state population in the $F=3$ ground level just before imaging is equally distributed among the $m_F$ states.

\section{Observations}

While the largest electric field that we were able to verify using \emph{in situ} dc Stark measurements of a laser cooled Rb cloud was 18~kV/cm, we were not limited by the breakdown of the plates.  In particular, we achieved 18~kV/cm by applying $\pm 30$~kV to the plates, and we have verified that no arcing occurs at applied voltages of up to $\pm 60$~kV.  Rather, our measurements were limited by a large loss of atoms from the MOT observed at applied fields above 15~kV/cm.  In addition to this loss, we also observed the decay of the electric field strength over time while the plates were at high voltage as well as a residual field that remained after the plates were grounded.  The effects of the decay and persistent residual fields could be mitigated by only leaving the plates on for a short time and by subsequently energizing the plates with the opposite polarity.

In order to study these effects in more detail, we applied an electric field in six different configurations. This was accomplished by using the switching network to apply either a negative or positive bias to one plate while keeping the other grounded, or to apply a positive (negative) bias to one plate and the opposite bias voltage to the other. The direction of the electric field was always oriented vertically, either pointing upwards or downwards depending on the bias voltage of the plates. We discuss below the results of these tests with the aim of guiding future designs and the usage of similar electric field plates.

%\begin{figure}[h!]
%\includegraphics[width=.45\textwidth]{plate_polarity.png}
%\caption{The six different plate configurations that were used during the testing of the electric field plates.}
%\label{fig:plate_configurations}
%\end{figure}

\subsection{Electric field shielding} 

\begin{figure}[h!]
\includegraphics[width=.5\textwidth]{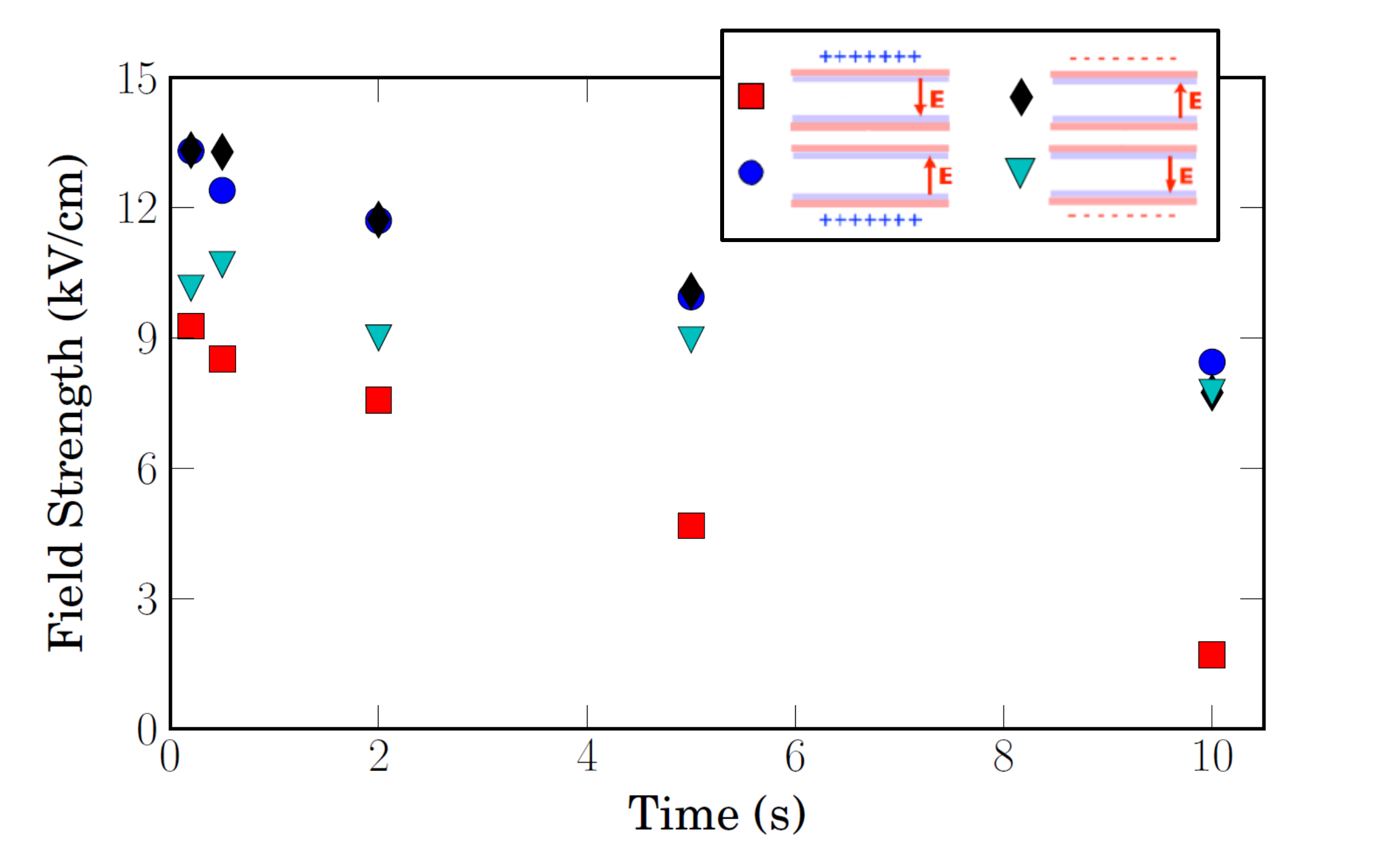}
\caption{(color online) The measured electric field strength at the center of the cell versus the time the plates are kept energized. In all four cases, one plate is biased at either $\pm 40~$kV, while the other plate is grounded. The low initial value and rapid reduction in the voltage occurs when the top plate is biased with $+40$~kV (red squares).  We note that the top plate is close to the metal support of the beam shield, and this behavior suggests that field emission of electrons is responsible for the rapid build-up of shielding charges. In the other cases, where the field emission rate of electrons should be lower, we observe a slower and less complete shielding of field at the center of the cell.}
\label{fig:strengh_vs_time}
\end{figure}

Initially, we studied the reduction of the electric field due to shielding as a function of the time the plates were connected to the high voltage supplies.
We first grounded one plate and applied either a positive or negative bias voltage of 40~kV to the other plate.  The behavior shown in Fig.~\ref{fig:strengh_vs_time} suggests that the applied voltage is shielded by the movement of free charges and the number of these charges, and thus the extent of the shielding, is influenced by the polarity of the bias voltage as well as the choice of which plate is energized and which is grounded.  Based on the data, we believe that the free charges responsible for this shielding behavior are primarily electrons produced by field emission of grounded metal parts within the chamber.  For sufficiently high voltages, we note that these free electrons can be accelerated to energies high enough to produce, upon collision with the cell walls, secondary positive ions that can also migrate and lead to shielding.

Fig.~\ref{fig:strengh_vs_time} shows the electric field strength measured some time after the plates were connected to the high voltage supply.  Each data point is an average of four measurements.  In order to minimize the effects of accumulated charges from previous measurements, the plates were run with the opposite polarity between each measurement, and we waited approximately 20 minutes before starting a data run for the next time in the series.

When the upper plate is positively biased, the shielding is the most rapid and dramatic.  This large and rapid shielding is in contrast to the moderate and slower shielding in all other configurations even when the lower plate is positively biased (with the upper plate grounded).  We believe that this asymmetry is due to the atomic beam shield (see Fig.~\ref{fig:TrappingLightStarkShift}) which has a metal support rod that runs along the top of the cell \cite{PhysRevA.88.023624}.  This support rod is made from stainless steel and is grounded by virtue of its contact with a CF flange that also supports the lithium oven.  Moreover, this rod was within 1~cm from the conductive layer inside the top electrode.  A large positive voltage applied near the grounded support rod is expected to lead to a much larger electron field emission than an equally large negative voltage.

We also observe shielding in the other configurations including those with a negative bias.  Any electrons on the surface or shallowly embedded inside the quartz cell wall will be accelerated away towards the grounded support rod or towards the opposite grounded plate when a negative bias is applied to a plate.  These electrons will, in doing so, acquire a considerable kinetic energy and could produce secondary positive ions when they impact the opposite cell wall or grounded support rod.  These positive ions will be accelerated back towards the negative plate and will also create secondary electrons when they impact the cell wall.  We believe that this process is what leads to more and more electric field shielding for longer ``on times", and we note that, based on the shielding behavior, the rates of secondary ion and electron production appear to be similar for the other three configurations.  In summary, the shielding behavior appears to be distinct in the case when the upper plate is positively biased and field emission of electrons from the grounded support rod is most probable.

\begin{figure}[h!]
\includegraphics[width=.45\textwidth]{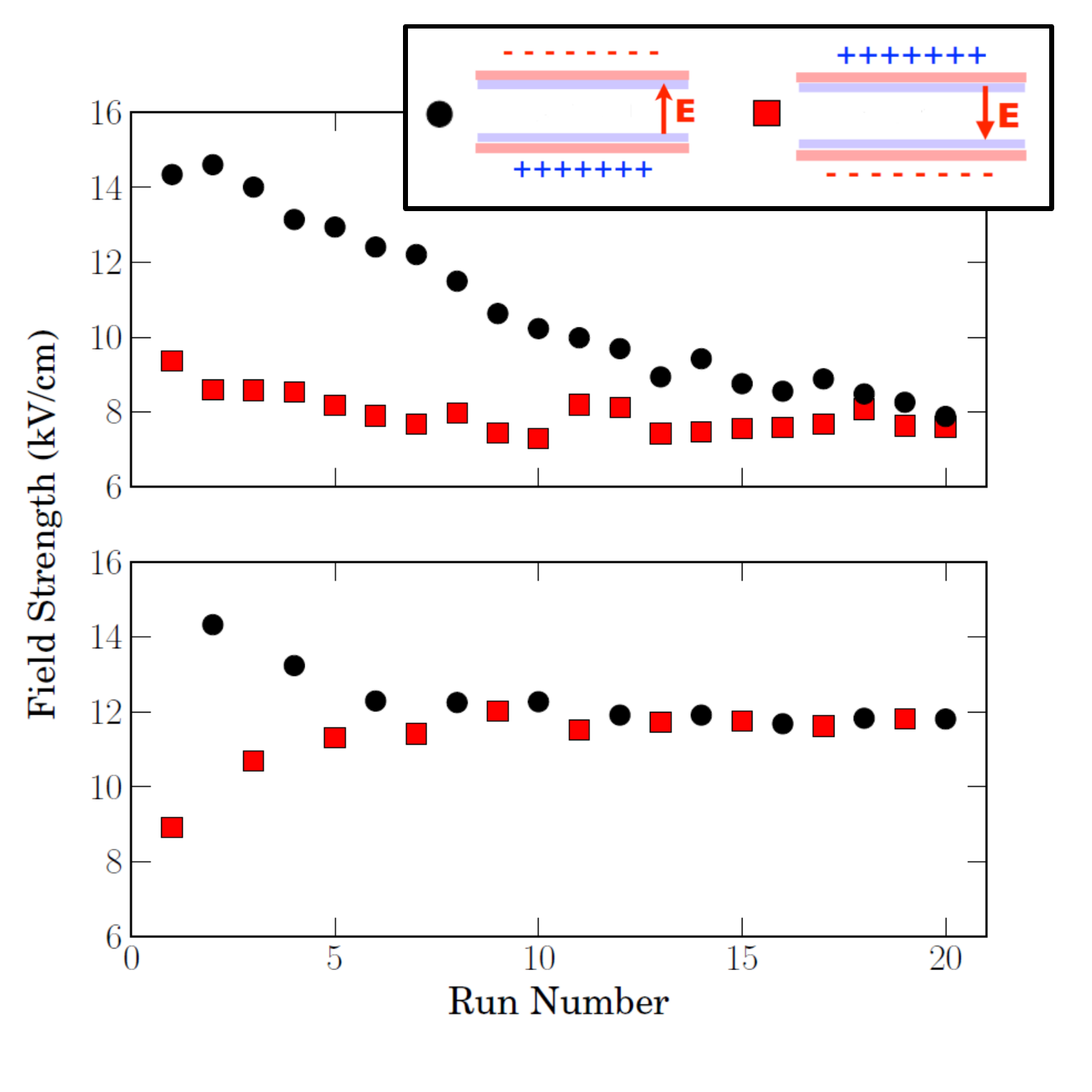}
\caption{(color online) The effects of shielding on the electric field strength over sequential experimental runs. The field is either kept in the same polarity for each run (top panel) or the polarity is switched between each run (bottom panel). In each run, the plates were energized for 500~ms before the field strength was measured. There is approximately 20~s between each run. The shielding observed in the top panel can be partially mitigated by alternating the polarity of the field between sequential runs.}
\label{fig:FieldDecay}
\end{figure}

The electric field plates were also operated with a positive bias on one plate and a negative bias on the other with a total potential difference of 40~kV.  In this case, the expected electric field at the center of the cell was 15~kV/cm.  To simulate using the plates as part of an experiment that required multiple repeated runs, we measured the field strength as a function of the run number.  During each experimental cycle, the plates were energized for 500~ms, and the time between each run was approximately 20~s. The top panel of Fig.~\ref{fig:FieldDecay} shows the measured field strength after each run, and it is evident that some of the shielding charges that are produced while the plates are on linger in the cell region during the time the plates are grounded leading to more and more free charges available for shielding and lower measured fields for subsequent runs.  The configuration with the upper plate positively biased (red squares) leads to a much larger number of shielding charges such that after the first run, with the plates on for just 500~ms, the field is already shielded to a value below 10~kV/cm. This is consistent with our observations in the case of a single positively biased top plate, shown in 
Fig.~\ref{fig:strengh_vs_time}.  By the 20th shot, the measured field in this configuration was only 7.5~kV/cm - approximately half the expected field.

When the upper plate is negatively biased and the lower plate positively biased (black circles), we observe, for the first few runs, a field magnitude very close to the expected field for the potential difference between the plates.  However, with each subsequent run, the measured field is less as a result of the lingering of the shielding charges during the off time of the plates.

It is neither helpful to have an electric field that changes in strength during the course of an experiment, nor to have a field that is weaker then theoretical achievable value due to a large amount of shielding. We attempted to fix both problems by flipping the field direction between each experimental run. As shown in the bottom panel of Fig.~\ref{fig:FieldDecay}, we found that after a few polarity changes, the field strength reached a steady state value of 12~kV/cm. This is larger than the maximally shielded value, though not as large as the expected field of approximately 15~kV/cm.

\subsection{Residual fields}

Not only did free charges within the chamber act to shield the electric field during the time in which the plates were energized, but they also resulted in a residual electric field that persisted within the vacuum cell after the electrodes were grounded.

As in the case of field shielding, the residual field was largest when a large positive bias was applied to the top plate. For example, 250~ms after a $40~$kV positive bias voltage was applied to the top plate (while keeping the bottom plate grounded), the measured electric field strength in the center of the cell was 9~kV/cm, and 250~ms later after the top plate was grounded, a 5~kV/cm residual electric field was observed in the center of the cell.  Although we did not verify this, we assume that the residual electric field was oriented antiparallel to the expected field direction and was equal to the difference between the expected field magnitude and the measured field magnitude for each configuration.

These observations suggest that the shielding charges that are created and accumulate while the plates are energized actually become embedded in the quartz cell walls such that they do not immediately leave when the plates are grounded.  Nevertheless, we observed that it was possible to remove the residual field by running the electric field plates in the opposite polarity. This was the primary motivation for the construction of the high voltage switching network and appears to be critical when free charges (or, more importantly, a source of free charges) exist near the electric field region.

\subsection{Trap loss}

In addition to the shielding effects, we also observed a loss of Rb atoms held in the MOT when large voltages were applied to the plates.  This observed atom loss was the primary technical limitation preventing us from measuring fields with a strength greater than 18~kV/cm.  In addition, we also observed a loss of Li atoms held in the MOT and ODT. Since our interest was in applying large electric fields to ultra-cold atoms held in an optical dipole trap (ODT), we studied the loss of $^6$Li from an ODT.

\begin{figure}[h!]
\includegraphics[width=.45\textwidth]{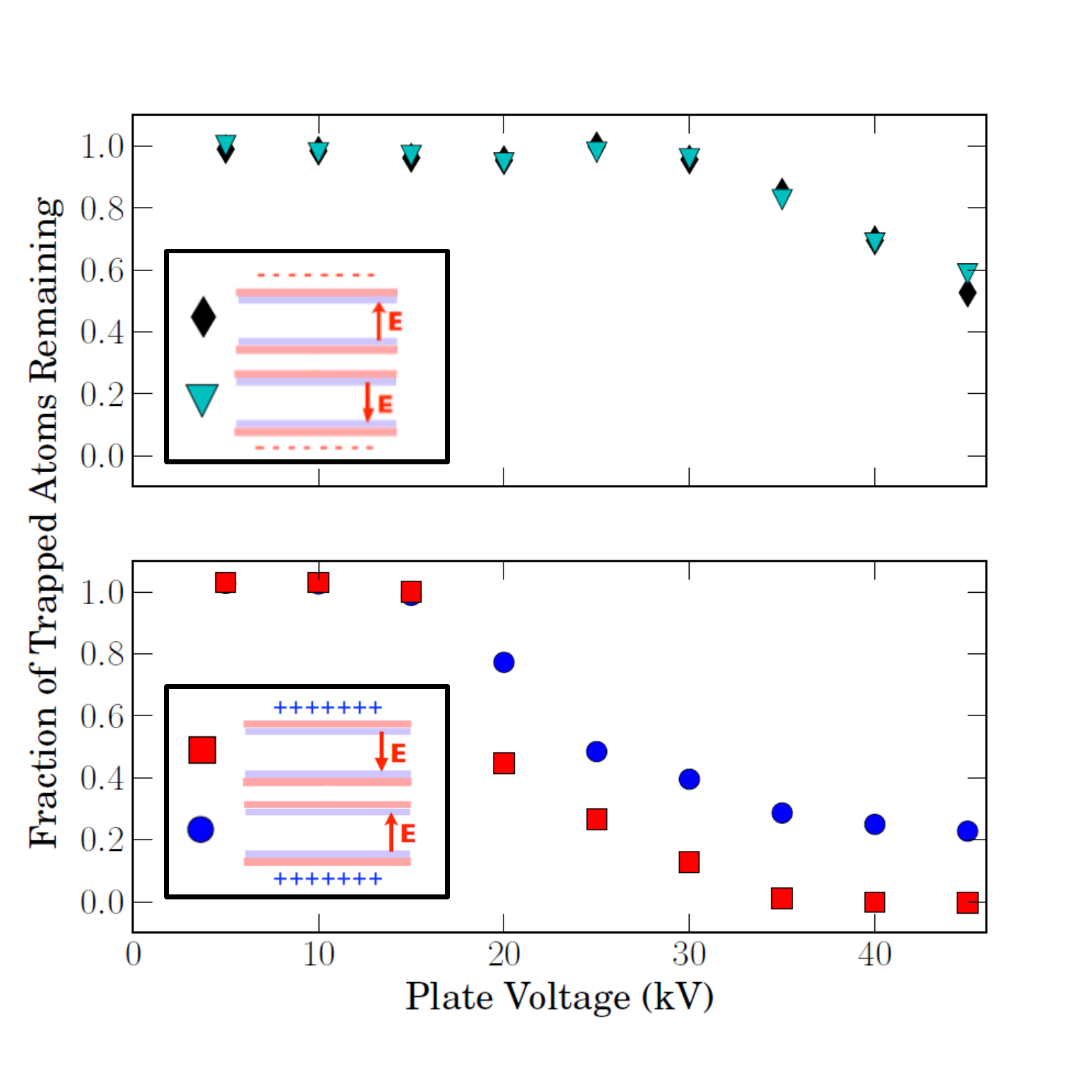}
\caption{(Color online) Loss of $^6$Li atoms from the ODT as a function of the negative (top panel) and positive (bottom panel) plate voltage with the second plate grounded.  The atom loss is much stronger when a positive bias is used (especially when the top plate is positively biased) due to strong field emission of electrons of the support arm of the beam shield. The weaker loss when a negative bias is used is attributed to the acceleration of free electrons that already exist within the vacuum or that are partially embedded and subsequently ejected from the quartz cell during the on time of the plate bias.}
\label{fig:ODTLoss}
\end{figure}

As Fig.~\ref{fig:ODTLoss} shows, we found that the $^6$Li atom loss was larger, and became observable at a smaller plate voltages, when using a positive bias voltage versus a negative bias voltage.  Moreover, the loss due to the positive bias was larger when the upper plate was charged and the lower plate was grounded.  This asymmetry is consistent with our field shielding observations where the accumulation of shielding charges is largest when the upper plate was positively biased.  In the case of a negatively biased plate (with the other plate grounded), there is almost no difference in the atom loss associated with the two configurations.  Along with the observations of field shielding, this suggests that collisions with accelerated electrons and secondary ions within the chamber are responsible for the atom loss from the ODT.  

We note that we did not see observable loss from the $^6$Li MOT when only one plate was charged even up to 40~kV (corresponding to an expected electric field of 15~kV/cm). However, we did observe $^6$Li atom loss from the MOT when both plates where charged (one with a negative bias, and the other with a positive bias) such that the total potential difference between the plates was larger than 50~kV. These observations indicate that the loss rate (similar to that for elastic collisions with neutral atoms \cite{PhysRevA.84.022708}) is smaller from a deeper trap (for the same electric field plate potential difference), and suggests that not all collisions ionize atoms (which would result in trap loss regardless of the trap depth).

\subsection{Tests with a smaller electric field plate}

In our prototyping stage, some initial tests were conducted with smaller electric field plates.  In particular, the conductive layer was constructed out of a much smaller ITO slide, measuring just under 3~cm $\times$ 3~cm.  Using these smaller plates, even at similar applied voltages and observed fields up to 15~kV/cm, we did not observe field shielding or the trap loss effects described above. As both plates were centered on the quartz cell, we attribute this difference in behavior to the larger distance (more than 4~cm) between the plates and the beam shield support rod. We believe that this observation re-enforces the hypothesis that field emission of electrons from the top of the beam shield is largely responsible for the fast shielding of the electric field and the atoms loss.  Therefore, in order to achieve large electric field strengths, commensurate with the potential difference applied across the plates, it is crucial to ensure that the region inside of the vacuum chamber near to the plates is free of any charge source (for example, grounded metal parts).

\section{Conclusion}

We have demonstrated a design for a transparent electrode for high electric field production using a buried ITO layer. The electrodes can be operated in air at standard pressure without suffering from electrical breakdown, even for fields which far exceed the dielectric breakdown of air.  We directly verify the production of electric fields up to 18~kV/cm. The maximum electric field we produce and verify is not limited by breakdown of the plates, but rather a large loss of atoms from the MOT at electric fields above 15~kV/cm due to collisions with accelerated free charges within the vacuum chamber. In addition, we observe an accumulation of charge which results in shielding of the electric field. We note that the shielding effect and atom loss is asymmetric and larger for a positive bias, which suggests that electron field emission from a beam shield inside the vacuum chamber is primarily responsible. While these observations suggest that using larger negative biases would be better, we found that it is technically challenging to work with voltages exceeding 60~kV. Therefore, large potential differences likely require the application of both a positive and negative voltage, and therefore it is critical to ensure that the vacuum chamber near the field region is free of any charge sources.

\begin{acknowledgments} 
The authors acknowledge financial support from the Natural Sciences and Engineering Research Council of Canada (NSERC / CRSNG), and the Canadian Foundation for Innovation (CFI). 

%We would also like to acknowledge Pavel Trochtchanovitch and Tony Mittertreiner along with the electronic shop in the Physics and Chemistry department at UBC for their help with the issues surrounding high voltage electronics and switching.  
\end{acknowledgments}

\bibliographystyle{osajnl}
\bibliography{RSI-ElecFieldPlate-V5_arx}
%\begin{thebibliography}{99}
%%% Do not include separate BibTeX files; if BibTeX is used,
%%% paste the output (contents of .bbl file) here.
%
%
%\end{thebibliography}

\end{document}